\documentclass[runningheads]{llncs}
\usepackage{multirow}
\usepackage[T1]{fontenc}
\usepackage{booktabs} 
\usepackage{hyperref}
\usepackage{graphicx}
\usepackage{adjustbox}
\usepackage{threeparttable}
\begin{document}
\title{Reconstructing Context}
\subtitle{Evaluating Advanced Chunking Strategies for Retrieval-Augmented Generation}

\author{
Carlo Merola\orcidID{0009-0000-1088-1495} \and
Jaspinder Singh\orcidID{0009-0000-5147-1249}
\thanks{Equal contribution.} } 
%\author{Anonymous authors}
%
\authorrunning{J. Singh and C. Merola}
\institute{Department of Computer Science and Engineering, University of Bologna
\email{carlo.merola@studio.unibo.it, jaspinder.singh@studio.unibo.it}}
\maketitle              % typeset the header of the contribution
\begin{abstract} Retrieval-augmented generation (RAG) has become a transformative approach for enhancing large language models (LLMs) by grounding their outputs in external knowledge sources. Yet, a critical question persists: how can vast volumes of external knowledge be managed effectively within the input constraints of LLMs? Traditional methods address this by chunking external documents into smaller, fixed-size segments. While this approach alleviates input limitations, it often fragments context, resulting in incomplete retrieval and diminished coherence in generation.
To overcome these shortcomings, two advanced techniques—late chunking and contextual retrieval—have been introduced, both aiming to preserve global context. Despite their potential, their comparative strengths and limitations remain unclear. This study presents a rigorous analysis of late chunking and contextual retrieval, evaluating their effectiveness and efficiency in optimizing RAG systems.
Our results indicate that contextual retrieval preserves semantic coherence more effectively but requires greater computational resources. In contrast, late chunking offers higher efficiency but tends to sacrifice relevance and completeness.
\end{abstract}
\section{Introduction}
Retrieval Augmented Generation (RAG) is a transformative approach that enhances the capabilities of large language models (LLMs) by integrating external information retrieval directly into the text generation process. This method allows LLMs to dynamically access and utilize relevant external knowledge, significantly improving their ability to generate accurate, contextually grounded, and informative responses.
Unlike static LLMs that rely solely on pre-trained data, RAG-enabled models can access up-to-date and domain-specific information. This dynamic integration ensures that the generated content remains both relevant and accurate, even in rapidly evolving or specialized fields.

RAG models combine two key components: a retrieval mechanism and a generative model. The retrieval mechanism fetches relevant documents or data from a large corpus, while the generative model synthesizes this information into coherent, contextually enriched answers. This synergy enhances performance in knowledge-intensive natural language processing (NLP) tasks, enabling models to produce well-informed responses grounded in the retrieved data.

%In recent years, the adoption of RAG has demonstrated promising results, particularly in tasks requiring deep knowledge integration. By bridging retrieval and generation, RAG represents a significant step forward in the evolution of NLP systems, ensuring that AI-generated content remains both insightful and up to date.

{\textbf{\textit{The Context Dilemma in Classic RAG:}}}
Managing extensive external documents poses significant issues in RAG systems. Despite advancements, many LLMs are limited to processing a few thousand tokens. Although some models have achieved context windows up to millions of tokens \cite{DBLP:conf/icml/DingZZXSX0Y24}, these are exceptions rather than the norm. Moreover, research indicates that LLMs may exhibit positional bias, performing better with information at the beginning of a document and struggling with content located in the middle or toward the end \cite{hsieh2024middlecalibratingpositionalattention,lu2024insightsllmlongcontextfailures}. This issue is exacerbated when retrieval fails to prioritize relevant information properly.
Thus, documents are often divided into smaller segments or "chunks" before embedding and retrieval. However, this chunking process can disrupt semantic coherence, leading to:
\begin{itemize}
    \item \textit{Loss of Context:} dividing documents without considering semantic boundaries can result in chunks that lack sufficient context, impairing the model's ability to generate accurate and coherent responses.
    \item \textit{Incomplete Information Retrieval:} important information split across chunks may not be effectively retrieved or integrated.
\end{itemize}

To address these issues, we analyse and compare two recent techniques---contextual retrieval\footnote{\url{https://www.anthropic.com/news/contextual-retrieval}} and late chunking \cite{günther2024latechunkingcontextualchunk}---within a unified setup, evaluating their strengths and limitations in tackling challenges like context loss and incomplete information retrieval. Contextual retrieval preserves coherence by prepending LLM-generated context to chunks, while late chunking embeds entire documents to retain global context before segmenting.

Our study rigorously assesses their impact on generation performance in question-answering tasks, finding that neither technique offers a definitive solution. This work highlights the trade-offs between these methods and provides practical guidance for optimizing RAG systems.

To further support the community, we release all code, prompts, and data under the permissive MIT license, enabling full reproducibility and empowering practitioners to adapt and extend our work.\footnote{\url{https://github.com/disi-unibo-nlp/rag-when-how-chunk}}

\section{Related Work}

\paragraph{Classic RAG.}
A standard RAG workflow involves four main stages: document segmentation, chunk embedding, indexing, and retrieval. During segmentation, documents are divided into manageable chunks. These chunks are then transformed into vector representations using encoder models, often normalized to ensure unit magnitudes. The resulting embeddings are stored in indexed vector databases, enabling efficient approximate similarity searches. Retrieval involves comparing query embeddings with the stored embeddings using metrics such as cosine similarity or Euclidean distance, which identify the most relevant chunks.
Seminal works like \cite{DBLP:conf/nips/LewisPPPKGKLYR020} and \cite{karpukhin2020densepassageretrievalopendomain} have demonstrated the effectiveness of RAG in tasks such as open-domain question answering. More recent studies, including \cite{gao2024retrievalaugmentedgenerationlargelanguage}, have introduced advancements in scalability and embedding techniques, further establishing RAG as a foundational framework for knowledge-intensive applications.

\paragraph{Document Segmentation.}
Document segmentation is essential for processing long texts in RAG workflows, with methods ranging from \textit{fixed-size segmentation} \cite{gao2024retrievalaugmentedgenerationlargelanguage} to more adaptive techniques like \textit{semantic segmentation},\footnote{\url{https://docs.llamaindex.ai/en/stable/examples/node\_parsers/semantic\_chunking/}} which detect semantic breakpoints based on shifts in meaning.
Recent advancements include \textit{supervised segmentation models} \cite{koshorek2018textsegmentationsupervisedlearning,jina2023finding} and \textit{segment-then-predict models}, trained end-to-end without explicit labels to optimize chunking for downstream task performance \cite{MORO2023126356}.
In 2024, \textit{late chunking} and \textit{contextual retrieval} introduced novel paradigms. Both techniques have proven effective in retrieval benchmarks but remain largely untested in integrated RAG workflows. Despite several RAG surveys \cite{gao2024retrievalaugmentedgenerationlargelanguage,fan2024surveyragmeetingllms,unknown}, no prior work has compared these methods within a comprehensive evaluation framework. This study addresses this gap by holistically analyzing late chunking and contextual retrieval, offering actionable insights into their relative strengths and trade-offs.

\section{Methodology}
To guide our study, we define the following research questions (RQs), aimed at evaluating different strategies for chunking and retrieval in RAG systems:
\begin{itemize}
    \item \textbf{RQ\#1}: Compares the effectiveness of \textbf{early versus late chunking} strategies, utilizing \textbf{different text segmenters} and embedding models to evaluate their impact on retrieval accuracy and downstream performance in RAG systems.
    \item \textbf{RQ\#2}:  Compares the effectiveness of \textbf{contextual retrieval versus traditional early chunking} strategies,  utilizing \textbf{different text segmenters} and embedding models to evaluate their impact on retrieval accuracy and downstream performance in RAG systems.
\end{itemize}

\subsection{RQ\#1: Early or Late Chunking?}
\begin{figure}[!tb]
  \centering
  \includegraphics[width=0.75\textwidth]{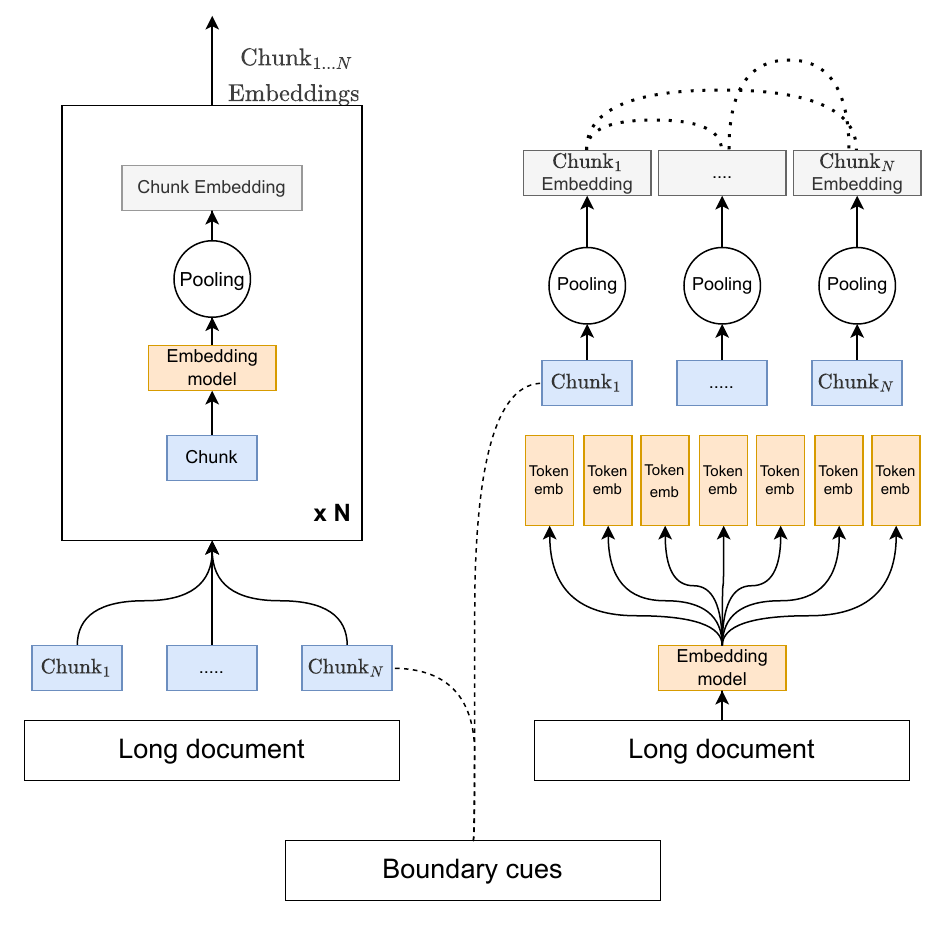}\label{latevse}
  \caption{Comparison of early chunking (left) and late chunking (right) approaches for processing long documents. In early chunking, the document is divided into chunks before embedding, with each chunk processed independently by the embedding model and then pooled. In contrast, late chunking processes the entire document to generate token embeddings first, using boundary cues to create chunk embeddings, which are subsequently pooled.}
\end{figure}

In this workflow, the main architectural modification compared to the standard RAG lies in the document embedding process Figure \ref{latevse}. Specifically, we experiment with various embedding models to encode document chunks, tailoring them to align with the early and late chunking strategies under evaluation. This adjustment allows us to explore how different embedding techniques influence the retrieval quality and, subsequently, the overall performance of the RAG system. Additionally, we test dynamic segmenting models to further refine the chunking process, providing an adaptive mechanism that adjusts chunk sizes based on content characteristics. By evaluating the impact of these dynamic segmenting models, we aim to improve the overall retrieval efficiency and response generation within the RAG framework.

\paragraph{Early Chunking.}
Documents are segmented into text chunks, and each chunk is processed by the embedding model. The model generates token-level embeddings for each chunk, which are subsequently aggregated using mean pooling to produce a single embedding per chunk.

\paragraph{Late Chunking.}
Late chunking \cite{günther2024latechunkingcontextualchunk} defers the chunking process. As shown in Figure ~\ref{latevse}, instead of segmenting the document initially, the entire document is first embedded at the token level. The resulting token embeddings are then segmented into chunks, and mean pooling is applied to each chunk to generate the final embeddings. This approach preserves the full contextual information within the document, potentially leading to superior results across various retrieval tasks. It is adaptable to a wide range of long-context embedding models and can be implemented without additional training.
The two approaches are tested with different embedding models. 

\subsection{RQ\#2: Early or Contextual Chunking?}
\label{contextual_methodology}

In this workflow, traditional retrieval is compared to Contextual Retrieval with Rank Fusion technique. This has been introduced by Anthropic in September 2024.\footnote{\url{https://www.anthropic.com/news/contextual-retrieval}}
Three steps are added to the Traditional RAG process: Contextualization, Rank Fusion, Reraking.

\paragraph{Contextualization.}

\begin{figure}[!tb]
  \centering
  \includegraphics[width=0.75\textwidth]{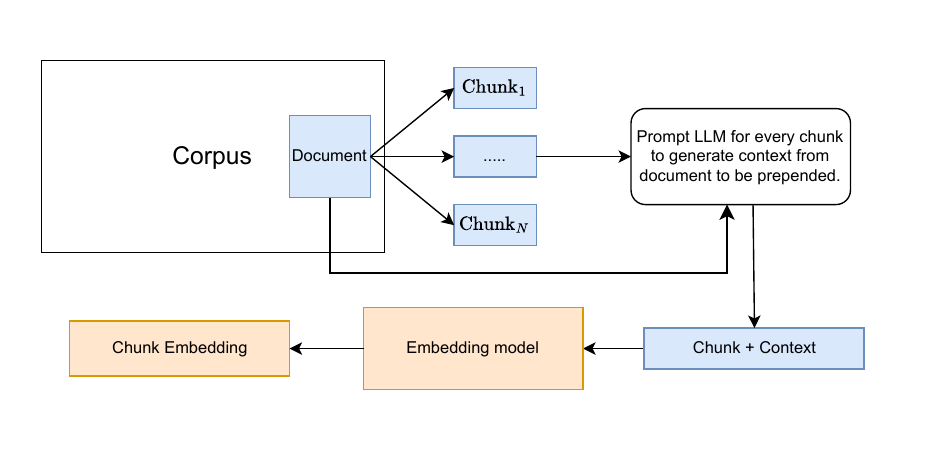}
  \label{fig:ContextualRetrieval}
  \caption{Contextualization of each chunk is performed prior to embedding. The document is divided into chunks, and a prompt is used to query an LLM to generate contextual information from the document for each chunk. The context is prepended to the chunk, which is then processed by the embedding model to produce the final chunk embedding.}
\end{figure}

After document segmentation, each chunk is enriched with additional context from the entire document, ensuring that even when segmented, each piece retains a broader understanding of the content (Fig. \ref{fig:ContextualRetrieval}).
In fact, when documents are split into smaller chunks, it might arise the problem where individual chunks lack sufficient context.
For example, a chunk might contain the text: "The company’s revenue grew by 3\% over the previous quarter." However, this chunk on its own does not specify which company it is referring to or the relevant time period, making it difficult to retrieve the right information or use the information effectively.
Contextualization improves the relevance and accuracy of retrieved information by maintaining contextual integrity.

\paragraph{Rank Fusion.}
In our methodology, we employ a rank fusion strategy that integrates dense embeddings with sparse embeddings of BM25 \cite{article_bm25} to improve retrieval performance. Although embedding models adeptly capture semantic relationships, they may overlook exact matches, which is particularly useful for unique identifiers or technical terms. BM25 uses a ranking function that builds upon Term Frequency-Inverse Document Frequency (TF-IDF), addressing this limitation by emphasizing precise lexical matches.
To combine the strengths of both approaches, we conduct searches across both dense embedding vectors and BM25 sparse embedding vectors generated from both the chunk and its generated context. Initially, the assigned relative importance in the search for the two vector fields has been set to be of equal intensity, resulting in lowering the scoring results in the retrieval evaluation. For this reason we use a weighting strategy assigning higher weights to dense vector fields, emphasizing them more in the final ranking. While different weight parameters have been tested, the final decision has been to define a ratio of importance 4:1 assigning weight 1 for the dense embedding vectors and 0.25 for the BM25 sparse embedding vectors. This ratio reflects Anthropic weight assignment.\footnote{\url{https://github.com/anthropics/anthropic-cookbook/blob/main/skills/contextual-embeddings/guide.ipynb}} By applying these weights, dense embeddings are emphasized more heavily, while still incorporating the contributions of the BM25 sparse embeddings.
This weighted rank fusion approach leverages the complementary strengths of semantic embeddings and lexical matching, aiming to improve the accuracy and relevance of the retrieved results.

\paragraph{Reranking Step.} 
To boost and obtain consistent improved performance, the retrieval and ranking stages have been separated. This two-stage approach improves efficiency and effectiveness by leveraging the strengths of different models at each stage.
After retrieving the initial set of document chunks, we implement an additional reranking step to enhance the relevance of the results to the user's query. This process involves reassessing and reordering the retrieved chunks based on their pertinence, prioritising the most relevant information.
The reranker operates by evaluating the semantic similarity and contextual relevance between the query and each retrieved chunk. It assigns a relevance score to each chunk, and the chunks are then reordered based on these scores, with higher-scoring chunks placed at the top of the list presented to the user. This method ensures that the most pertinent information is readily accessible, improving the overall effectiveness of the retrieval system.
Implementing this reranking step addresses potential limitations of the initial retrieval process, such as the inclusion of less relevant chunks or the misordering of pertinent information.

\section{Experimental Setup}
\label{sec:init_setup}
This study has focused on testing these techniques with open-source models. A particular focus was also given to resource usage, for real-world scenarios that lead towards the choice of the LLM for the question answering task to Microsoft's \texttt{Phi-3.5-mini-instruct},\footnote{\url{https://huggingface.co/microsoft/Phi-3.5-mini-instruct}} quantized to 4 bits. \cite{abdin2024phi3technicalreporthighly} This language model is designed to operate efficiently in memory- and compute-constrained environments which is a crucial aspect of our work. The same language model has been used also to generate additional context to prepend to each chunk in the Contextual Retrieval Setup.

For what regards the embedding models these ones have been tested: \texttt{Jina V3} \cite{sturua2024jinaembeddingsv3multilingualembeddingstask}, \texttt{Jina Colbert V2} \cite{günther2024jinaembeddings28192token}, \texttt{Stella V5} and \texttt{BGE-M3} \cite{chen2024bgem3embeddingmultilingualmultifunctionality}, all present in the \texttt{MTEB} \cite{muennighoff2023mtebmassivetextembedding} leaderboard (see \autoref{tab:model_paramters} for more details).
\paragraph{Dataset and Hardware.} There are severe limitations in current datasets available for RAG systems evaluation. Many don't include together the labels for retrieval quality evaluation and answers labels for the quality of the generation in a question answering system. In our system initial Retrieval performance has been tested on the \texttt{NFCorpus} \cite{inproceedicngs} dataset, while the subsequent Generation performance in question answering has been conducted over \texttt{MSMarco}\cite{bajaj2018msmarcohumangenerated}.

Another important note for Contextual Retrieval (\texttt{RQ\#2}). The \texttt{NFCorpus} dataset is characterised by a long average document length. Here appear the first limitations of the Contextual Retrieval approach to RAG. For the intrinsic nature of this approach, the segmented chunks are enhanced with a generated context taken from the document, prompting an LLM for the task, leveraging the new advents of Instruction Learning. Chunks and documents are passed together in a formatted prompt to the model. When a document reaches long lengths, the VRAM of the GPU gets filled up quickly. For chunk contextualization, around 20GB of VRAM use can be reached, limiting batch dimensions for generation and slowing down the times needed for effective chunk contextualization.

In our experimental setup, we utilized an Nvidia RTX 4090 with 24GB of VRAM. Due to GPU memory constraints, we employed a subset of the dataset, corresponding to 20\% of the entire \texttt{NFCorpus} for \texttt{RQ\#2}, while the full dataset was used for \texttt{RQ\#1} workflow.

For datasets such as \texttt{MsMarco}, which include only passage texts rather than full documents, the system operates within a more constrained context for generating responses. This limitation arises because passages are typically shorter segments of text, providing less information for contextual understanding. As a result in \texttt{RQ\#2}, the system's ability to generate contextually relevant and comprehensive responses can be affected by the brevity of the input text, potentially impacting the quality and depth of the generated content.

In \texttt{RQ\#1}, the evaluation was conducted on the first 1,000 queries and approximately 5,000 documents/passages. For \texttt{RQ\#2}, due to the significant computational requirements and hardware limitations, the experiments were restricted to 50 queries and around 300 documents.

\subsection{Embedding generation}\label{cak}

\paragraph{Common to both RQs.}
To generate embeddings for our experiments, we utilized different embedding models, as detailed previously (see section \ref{sec:init_setup}).
Each segmentation model approach outlined was paired with an appropriate embedding model to evaluate its influence on downstream tasks.

For fixed-size segmentation, we divided the text into equal-sized chunks with a predefined length of 512 characters. This approach ensures uniform chunk sizes, simplifying processing and offering a baseline for comparison with more adaptive methods.

For semantic segmentation, we used the \texttt{Jina-Segmenter API},\footnote{\url{https://jina.ai/embeddings/}} which dynamically adjusts chunk boundaries based on the semantic structure of the text. This ensures that the segments capture meaningful content, improving the quality of embeddings generated.

All the generated embeddings were normalized to unit vectors, facilitating cosine similarity computations during the retrieval phase and ensuring uniformity across experiments.

\paragraph{RQ\#1:} 
In addition to the mentioned segmentation approaches, for this workflow, dynamic segmentation was tested, testing two models to assess their performance. The first model, \texttt{simple-qwen-0.5}, is a straightforward solution designed to identify boundaries based on the structural elements of the document. Its simplicity makes it efficient for basic segmentation needs, offering a computationally lightweight approach.

The second model, \texttt{topic-qwen-0.5}, inspired by Chain-of-Thought reasoning, enhances segmentation by identifying topics within the text. By segmenting text based on topic boundaries, this approach captures richer semantic relationships, making it suitable for tasks requiring a deeper understanding of document content.

\begin{table}[!t]
    \centering
    \resizebox{\textwidth}{!}
    {
    \begin{tabular}{|l||c|c|c|c|c|c|}
    \hline
     Model & MTEB Rank & Model Size(M) & Memory(GB) & Embedding Dim & Max Token  \\ \hline
    Stella-V5  & 5 & 1,543 & 5.75 & 1,024 & 131,072 \\
    Jina-V3\cite{sturua2024jinaembeddingsv3multilingualembeddingstask}   & 53 & 572 & 2.13 & 1,024 & 8,194 \\
    Jina-V2 \cite{günther2024jinaembeddings28192token}   & 123 & 137 & 0.51  & 1,024 & 8,194  \\
    BGE-M3 \cite{chen2024bgem3embeddingmultilingualmultifunctionality} & 211 & 567 & 2.11 & 1,024 & 8,192 \\ \hline
    \end{tabular}
    }
    
\caption{Embedding models.}
\label{tab:model_paramters}
\end{table}

\paragraph{RQ\#2:}

In this workflow, for the ContextualRankFusion evaluation, contextualization of the chunks before the embedding is necessary.
To contextualize each chunk, we prompt Microsoft's LLM model \texttt{Phi3.5-mini-instruct} to generate a brief summary that situates the chunk within the overall document, formatting the prompt with the chunk and its relative original document.

\subsection{Retrieval Evaluation}

\paragraph{RQ\#2:}
In this workflow, specifically for the ContextualRankFusion retrieval, the document retrieval has been enhanced with two additional steps (see Section \ref{contextual_methodology}).
In the reranking step, Rank Fusion was allowed through \texttt{Milvus} Vector Database, which integrates with \texttt{BM25} through the \texttt{BM25EmbeddingFunction} class, enabling hybrid search across dense and sparse vector fields.\\
After retrieving the top documents, these are reordered through the reranker model \texttt{Jina Reranker V2 Base} model,\footnote{\url{https://huggingface.co/jinaai/jina-reranker-v2-base-multilingual}} that employes a cross-encoder architecture that processes each query-document pair individually, outputting a relevance score. This design enables the model to capture intricate semantic relationships between the query and the document before being given to the LLM.

\paragraph{Scorings.}
For both approaches in \texttt{RQ\#1} and \texttt{RQ\#2}, when querying the embedding database (generated in \ref{cak}), the output will be a ranked list of chunks, ordered from the most similar to the query to the least similar. We employ a straightforward aggregation strategy to transition from chunk-level rankings to document-level rankings. Specifically, for each document, we consider the score of its most significant chunk as the representative value for the entire document. This approach ensures that a document's relevance is determined by its most relevant chunk.

Once the document scores are determined, we generate a ranked list of documents based on these scores. From this ranking, we extract the top-k documents, focusing on the Top 5 or Top 10 documents, depending on the specific evaluation scenario. This final document ranking is then used to assess the effectiveness of the retrieval process.

This methodology highlights the importance of individual chunks in influencing the overall document ranking and ensures that highly relevant chunks directly impact the document's position in the final ranking.

\paragraph{Metrics.}
To evaluate the performance of our model, we utilize three key metrics: NDCG , MAP, and F1-score. Each metric serves a specific purpose in assessing different aspects of the results.
\textit{Normalized Discounted Cumulative Gain (NDCG):} It measures the usefulness of an item based on its position in the ranking, assigning higher weights to items appearing at the top of the list. By using NDCG, we aim to assess the relevance of predictions in a way that prioritizes higher-ranked items.
\textit{Mean Average Precision (MAP)}:
It calculates the mean of the Average Precision (AP) scores for all queries, where AP considers the precision at each relevant item in the ranked list.  With MAP, we aim to quantify how effectively the model retrieves relevant results across different scenarios.
\textit{F1-score:}
The F1-score, a harmonic mean of precision and recall, is employed to balance the trade-off between false positives and false negatives.

\subsection{Generation evaluation} 
Generation evaluation was assessed through the \texttt{MSMarco} dataset using a question answering task. While for the Late Chunking technique, the scoring on the generation respects the retrieval performance, for the Contextual Retrieval Setup, chunks are enriched with additional generated context from the document that influences the output generation of an LLM. Although some differences were measured in the scorings, they were not notable enough to assess a significant difference in generation performance.

\section{Results and Analysis}

\subsection{Traditional Retrieval Versus ContextualRankFusion Retrieval}
From the results in \autoref{tab:results_contextual}, and especially focusing the attention on the best performing embedding model   \texttt{Jina-V3}, we show that Fixed-Window Chunking versus Semantic Chunking techniques do not differ much in terms of performance or not at all, while the first one being far easier implementable and faster than the second one. A more important finding underlines in the Rank Fusion technique. This technique shows improved results especially when chunks are enriched with additional context from the document. In this way, \texttt{BM25} matches search terms in both segments and contexts, leading to very good results.
It is important to note that adding the final reranking step in the workflow is crucial to leverage this potential and see consistent improvements in the results.

\begin{table}[!tb]
    \centering
    \begin{adjustbox}{width=0.75\textwidth}
    \begin{tabular}{|c|c|c||c|c|c|c|c|c|}
        \hline
        \textbf{Model} &\textbf{CM  } & \textbf{\centering RM} & \textbf{\centering NDCG@5} & \textbf{\centering MAP@5}  & \textbf{\centering F1@5}  & \textbf{\centering NDCG@10} & \textbf{\centering MAP@10}  & \textbf{\centering F1@10} \\
        \hline
        \multirow{8}{*}{Jina-V3}    & FUC   & TR    &0.303  &0.137  &0.193  &0.291  &0.154  &0.191\\
                                    &       & RFR   &0.289  &0.130	&0.185	&0.288  &0.150  &0.193\\
        \cline{2-9}  
                                    & SUC   & TR    &0.307  &0.143	&0.197  &0.292  &0.159  &0.187\\
                                    &       & RFR   &0.295  &0.135	&0.194  &0.287  &0.152  &0.189\\
        \cline{2-9}  
                                    & FCC   & TR    &0.312  &0.144	&0.204	&0.295  &0.159	&0.190\\
                                    &       & RFR   &\textbf{0.317}  &\textbf{0.146}  &0.206  &0.308  &0.166 &0.202\\
        \cline{2-9}  
                                    & SCC   & TR    &0.305  &0.136	&0.197  &0.296  &0.155  &0.198\\
                                    &       & RFR   &0.317	&0.146	&\textbf{0.209}  &\textbf{0.309} 
                                    &\textbf{0.166}	 &\textbf{0.204}\\ 
        \hline
        \hline
        \hline
        \multirow{8}{*}{Jina-V2}    & FUC   & TR    &0.206  &0.084	&0.138  &0.202  &0.096  &0.137\\
                                    &       & RFR   &0.256  &0.119	&0.166  &0.251  &0.133  &0.161\\
        \cline{2-9}  
                                    & SUC   & TR    &0.231  &0.100	&0.152  &0.223  &0.112  &0.149\\
                                    &       & RFR   &0.274  &0.127  &0.179  &0.262  &0.140  &0.168\\
        \cline{2-9}  
                                    & FCC   & TR    &0.232  &0.098	&0.155	&0.219  &0.109  &0.143\\
                                    &       & RFR   &0.288  &0.130  &0.182	&0.274  &0.144	&0.173\\
        \cline{2-9}  
                                    & SCC   & TR    &0.231  &0.099	&0.156  &0.220  &0.110	&0.148\\
                                    &       & RFR   &0.297  &0.134  &0.191  &0.283  &0.148	&0.180\\
        \hline
        \hline
        \hline
        \multirow{8}{*}{BGE-M3}     & FUC   & TR    &0.017	&0.006  &0.015  &0.018  &0.007  &0.014\\
                                    &       & RFR   &0.032	&0.012  &0.018	&0.033  &0.012  &0.020\\
        \cline{2-9}  
                                    & SUC   & TR    &0.012	&0.003  &0.001	&0.012  &0.003  &0.011\\
                                    &       & RFR   &0.029  &0.008	&0.017	&0.026  &0.009  &0.018\\
        \cline{2-9}  
                                    & FCC   & TR    &0.007	&0.001	&0.003	&0.012  &0.003  &0.012\\
                                    &       & RFR   &0.040  &0.015  &0.026	&0.040  &0.016  &0.027\\
        \cline{2-9}  
                                    & SCC   & TR    &0.002  &0.001	&0.001	&0.006  &0.002  &0.007\\
                                    &       & RFR   &0.034  &0.014	&0.021	&0.030  &0.015  &0.019\\
        \hline
        \hline
        
    \end{tabular}
    \end{adjustbox}
    \caption{Comparative results on a subset of the \texttt{NFCorpus} dataset. 20\% of the whole shuffled dataset was taken, deleting labels of documents not present in the subset dataset for retrieval evaluation. Scorings will be higher on the whole dataset.}
    
    \begin{tablenotes}
    \item[1]\textbf{CM}: Chunking Methods (FUC: Fixed-Window Uncontextualized Chunks, SUC: Semantic Uncontextualized Chunks, FCC: Fixed-Window Contextualized Chunks, SCC: Semantic Contextualized Chunks).
    \item[2] \textbf{RM}: Retrieval Methods (TR: Traditional Retrieval, RFR: Rank Fusion with weighted strategy (1, 0.25) respectively for dense embedder models and BM25 embeddings -- additional Reranking step for RFR).
    \end{tablenotes}
    \label{tab:results_contextual}
    \end{table}

\subsection{Traditional Retrieval Versus Latechunking Retrieval}
Upon analyzing the results in \autoref{tab:res1}, we observe that the novel Late Chunking approach performs well in most cases when compared to the Early version. This indicates its potential as an effective retrieval strategy for many scenarios. However, it is important to note that Late Chunking does not consistently outperform the Early approach across all models and datasets. 

For instance, with the \texttt{BGE-M3} model applied to the \texttt{NFCorpus}, the Early version demonstrates superior performance, highlighting a case where the Late Chunking approach falls short. This observation is further confirmed through testing on the \texttt{MsMarco} dataset using the \texttt{Stella-V5} model (\autoref{tab:res2}), where once again the Early version outperforms the Late Chunking approach.

These findings suggest that while Late Chunking introduces promising improvements in certain contexts, its efficacy may vary depending on the dataset and model used, emphasizing the need for careful selection of retrieval strategies based on specific use cases.
\subsection{Latechunking Versus ContextualRankFusion Retrieval}
In \autoref{tab:res_cont_late} we compare the best results obtained for ContextualRankFusion with Latechunking on the same subset of \texttt{NFCorpus} in order to compare the two techniques. The embedding model used is \texttt{Jina-V3}, for Fixed-Window Chunks. ContextualRankFusion obtains better results overall.

\begin{table}[!tb]
    \centering
    \resizebox{\textwidth}{!}
    {
    \begin{tabular}{|l||c|c|c||c|c|c|c|c|c|}
    \hline
    \textbf{Model}       & \textbf{Chunk} & \textbf{Segm}  & \textbf{Length}   & \textbf{NDCG@5} & \textbf{MAP@5} &\textbf{F1@5} & \textbf{NDCG@10} & \textbf{MAP@10} &\textbf{F1@10}\\ \hline
     \multirow{2}{*}{Stella-V5}   
     & Early & Fix-size & 512  & 0.443 &  \textbf{0.137} & \textbf{0.226} &  \textbf{0.414} &  \textbf{0.161} & \textbf{0.247} \\
     & Late  & Fix-size & 512  & \textbf{0.445} & 0.133 & 0.225 & 0.410 & 0.158 & 0.242 \\ \hline
    \multirow{5}{*}{Jina-V3}   
        & Early & Fix-size  & 512   & 0.374 & 0.107 & 0.186 & 0.346 & 0.127 & 0.204 \\
        & Late  & Fix-size  & 512   & 0.380 & 0.103 & 0.185 & 0.354 & 0.125 & 0.210\\
        & Early & Jina-Sem  & -     & 0.377 & 0.111 & 0.192 & 0.353 & 0.130 & 0.210\\
        & Late  & sim-Qwen  & -     & 0.384 & 0.105 & 0.185 & 0.356 & 0.126 & 0.206\\
        & Late  & top-Qwen  & -     & 0.383 & 0.102 & 0.179 & 0.351 & 0.122 & 0.203\\ \hline
    \multirow{5}{*}{Jina-V2}        
        & Early & Fix-size  & 512   & 0.261 & 0.064 & 0.124 & 0.237 & 0.075 & 0.137\\
        & Late  & Fix-size  & 512   & 0.280 & 0.069 & 0.125 & 0.255 & 0.081 & 0.146\\        
        & Early & Jina-Sem  & -     & 0.294 & 0.079 & 0.144 & 0.269 & 0.092 & 0.158\\
        & Late  & sim-Qwen  & -     & 0.278 & 0.071 & 0.130 & 0.253 & 0.083 & 0.146\\
        & Late  & top-Qwen  & -     & 0.279 & 0.070 & 0.135 & 0.254 & 0.081 & 0.147\\ \hline
    \multirow{5}{*}{BGE-M3}         
        & Early & Fix-size  & 512   & 0.246 & 0.059 & 0.120 & 0.225 & 0.069 & 0.130\\
        & Late  & Fix-size  & 512   & 0.070 & 0.010 & 0.029 & 0.067 & 0.013 & 0.038\\
        & Early & Jina-Sem  & -     & 0.260 & 0.066 & 0.122 & 0.240 & 0.079 & 0.144\\
        & Late  & sim-Qwen  & -     & 0.091 & 0.015 & 0.038 & 0.081 & 0.018 & 0.045\\
        & Late  & top-Qwen  & -     & 0.110 & 0.019 & 0.044 & 0.097 & 0.022 & 0.048\\
    \hline
    \end{tabular}
    }
    \caption{EarlyVsLate Retriever comparison on \texttt{NFCorpus}. Bold values indicate the best performance for each metric}
    \label{tab:res1}
\end{table}

\begin{table}[!tb]
    \centering
    \resizebox{\textwidth}{!}
    {
    \begin{tabular}{|l||c|c|c||c|c|c|c|c|c|}
    \hline
    \textbf{Model}       & \textbf{Chunk} & \textbf{Segm}  & \textbf{Length}   & \textbf{NDCG@5} & \textbf{MAP@5} &\textbf{F1@5} & \textbf{NDCG@10} & \textbf{MAP@10} &\textbf{F1@10}\\ \hline
     \multirow{2}{*}{Stella-V5}   & Early & Fix-size & 512    & \textbf{0.630} &  \textbf{0.501} & \textbf{0.019}& \textbf{0.632} &  \textbf{0.502} &\textbf{0.011}\\
       & Late  & Fix-size & 512    & 0.503 & 0.340 & 0.018 & 0.505 & 0.341 & 0.010\\ 
    \hline
    \end{tabular}
    }
    \caption{EarlyVsLate Retriever comparison \texttt{MsMarco}. Bold values indicate the best performance for each metric.}
    \label{tab:res2}
\end{table}

\begin{table}[!tb]
    \centering
    %\resizebox{\textwidth }{!}{
    \begin{tabular}{|l|c|c|c|c|c|c|}
    \hline
    \textbf{Method}&\textbf{NDCG@5} & \textbf{MAP@5} &\textbf{F1@5} & \textbf{NDCG@10} & \textbf{MAP@10} &\textbf{F1@10}\\ 
    \hline
    Late        &0.309& 0.143& 0.202& 0.294& 0.160& 0.192\\
    Contextual  &\textbf{0.317}&\textbf{ 0.146}& \textbf{0.206}& \textbf{0.308}& \textbf{0.166}&\textbf{ 0.202}\\
    \hline
    \end{tabular}
    %}
    \caption{Latechunking (Late) comparison versus ContextualRankFusion (Contextual) best performances, on same \texttt{NFCorpus} dataset subset (20\% of the whole). Embedding Model: \texttt{Jina-V3}. Chunking Method: Fixed-Window Chunking.}
    \label{tab:res_cont_late}
\end{table}

\subsection{Dynamic segmenting models}
As shown in \autoref{tab:res1}, the performance of pipelines utilizing dynamic segmentation, such as with \texttt{Jina-V3}, is superior to other approaches. However, this improvement comes at the cost of increased computational requirements and longer processing times. Specifically, embedding the \texttt{NFCorpus} dataset entirely with our experimental setup with fixed-size or semantic segmenter takes approximately 30 minutes. In comparison, the \texttt{Simple-Qwen} model requires twice the time, while the \texttt{Topic-Qwen} model requires four times as long.

Another drawback of these models is their generative nature, which can lead to inconsistencies. They do not always produce the exact same wording for chunks, rendering them less reliable in certain scenarios.

\section{Conclusion}
While both approaches are effective solutions at mitigating the challenge of context-dilemma, maintaining context in document retrieval in certain scenarios, both cannot be considered definitive solutions to tackle the problem. Late chunking offers a more computationally efficient solution by leveraging the natural capabilities of embedding models. In contrast, contextual retrieval, with its reliance on LLMs for context augmentation and re-ranking, incurs higher computational expenses. It also notable that the type of document and it's length can affect the performances, together with the LLM chosen for the task, smaller and more efficient models performing worse.\\ This distinction is crucial for applications where computational resources are a significant consideration like in real-world scenarios.

\section*{Preprint Acknowledgment}
This preprint has not undergone peer review or any post-submission improvements or corrections. The Version of Record of this contribution will be available online in \emph{Second Workshop on Knowledge-Enhanced Information Retrieval, ECIR 2025}, 
Springer Lecture Notes in Computer Science, via Springer’s DOI link after publication.


\end{document}